# Key Economic and Environmental Perspectives on Sustainability in the ICT Sector

Hassan Hamdoun[1], Jafar A. Alzubi[2], Omar A. Alzubi[3], and Solomon Mangeni[4]

1. Managing/Researcher Director, dot.energy, Swansea ,UK, email: *h.hamdoun@dotenergyconsulting.com*
2. Assistant professor, Al-Balqa Applied University, Jordan, Email: *j.zubi@bau.edu.jo*
3. Assistant professor, Al-Balqa Applied University, Jordan, Email: *o.zubi@bau.edu.jo*
4. Researcher, College of Engineering, Swansea University, Email: *s.w.mangeni@swansea.ac.uk*

*Abstract*

Telecommunication networks have become as critical to the 21st century development as were railways, roads and canals, to the 19th Century developments and is now seen as enabler to a more sustained business, environment and society as a whole.

Still fascinating has been and is the exponential rate of growth in this industry. This is one sector where the next revolution is always just around the corner whether known or unknown. The telecoms industry is categorized by high rates of innovation in a rapidly changing technological environment. This in turn is associated with an immense range of sustainability concerns and challenges for the Telecoms service providers, the service users and the whole industry and it's far reaching influence on other industries. This paper discusses three key aspects of such challenges namely; the question of sustainable power/energy supply for the industry when the change is resulting in increasing energy and operational cost, the exploitation of technologies advancement for sustainability and their business and environmental benefits.

# I. INTRODUCTION

Telecommunications is one of the few industries that have evolved from being looked at as a technology competing with other industries to a technology that is enabling them. One can hardly think of any industry that can be run today without making good use of telecommunication technology. The popular question people ask one another goes like; how did we manage without these devices, phones, computers, tablets and their like?

This sector of industry is so pervasive that it not only advances at a furious rate but also impacts other industries and their growth at not only the same rate but orders of magnitude higher. The past few decades have witnessed a widespread diffusion of the World Wide Web (internet), followed by mobile phones and more recently the transition to using smart phones, smart meters and the eventual smart environments which are soon becoming a reality. These have led to an enormous increase in mobile data services. New technologies are now rapidly changing consumer requirements such as the need for continuous infrastructural modernization just to keep pace with these requirements. Heightening demand in maturing markets countered by rapid advancement in emerging markets and increased outsourcing of services present a number of sustainability challenges for Telecoms service providers [2].

So it is generally believed now that the internet, on which many services depend, is poised to become truly pervasive providing services anywhere, anytime connectivity to people objects and things [3].

# II. SUSTAINABILITY

Sustainability a word that has become widely used and become applicable in many disciplines of study, research and service delivery has also found its way into the telecommunications industry. There are many ways in which sustainability has been defined. The simplest and most central is: "the ability to sustain" or simply "the capacity to endure." These; the telecommunications industry has proven itself as an enduring technology. One definition that seems comprehensive says "sustainability refers to development which is capable of meeting the needs of the present without compromising the ability of future generations to meet their own needs" [4].

Logically, does a hungry, homeless or a person in a disaster stricken area ask for a telephone or internet? Obviously not. He or she will ask for food, water, shelter and medicine. However, meeting these basic needs naturally becomes the first priority but how quickly, with what efficiency? Are questions one may ask? Few governments in poor or disaster stricken countries might feel that telecommunication development in rural areas is something they can afford to worry about. However, timely provision of basic needs such as food, medicine and health care, and more particularly during disaster for rescue and relief operations depends heavily on the availability of the telecommunications infrastructure.

In a holistic way, the telecommunications industry attracts involvement and influence from all sectors of the economies. No single sector of growth can afford to ignore what is happening in the communications industry from politicians in the high rising city flats to peasants on a rural farm. One example that highlights this is the recognition of the 'Internet' as a basic human right by France and Estonia in 2011[5].

III. ENERGY & SUSTAINABILITY IN ICT

We live in a world today where 768g of $CO_2$ are required in order to generate £1 of Gross Domestic Product (GDP). For the world to be sustainable, £1 GDP should use ONLY 6g of $CO_2$. The operations of economic, social and political activities all contribute to this figure. Information Communication Technologies (ICT) is increasingly a major tool to aid economic, social and political growth of companies, organizations and governments [6].

Note here that there are two measures for emission measurements: Green House Gas (GHG) emissions and $CO_2$ emissions. $CO_2$ emissions constitute major gas emitted into the atmosphere but also portions of methane, nitrous oxide and fluorocarbons contribute to heating the atmosphere and included in GHG calculations. Globally, the ICT sector was estimated to have contributed up to 16% of GDP growth from 2002 to 2007 and the sector increased its share of GDP worldwide from 5.8 to 7.3%. The ICT sector's share of the global economy is predicted to jump further to 8.7% of GDP growth from 2007 to 2020[7, 8]. However what is the ICT contribution to the $CO_2$ emissions. The figure is growing year on year at 4% and will reach 2.3% of global GHG emissions by 2020 [9], with the telecommunications industry taking 2-3% of this share [10].

*A. ICT contributions to $Co_2$ emissions*

Figure 1 shows the ICT contribution to global carbon dioxide equivalent (CO2e)[11]. The ICT sector's own emissions are expected to increase, in a business as usual (BAU) scenario, from 0.53 billion tones (Gt) carbon dioxide equivalent (CO2e) in 2002 to 1.27 GtCO2e in 2020. However, the Global E-Sustainability Initiative (GeSi) report showed that ICT technologies and services would enable opportunities to reduce emissions seven times the size of the sector's own footprint, up to 9.1 GtCO2e, or 16.5% of total BAU emissions by 2020 [12].

As can be seen from Figure 1, the emission growth is increasing by 4.6% Compound Annual Growth Rate (CAGR) and 7.1% for voice and Data networks and Data centers respectively. This is compared to only 2.3% CAGR increase from end-user devices.

This paper aims to shed a light on the role of powering ICT in sustainability challenges address the need in improving power efficiency in telecommunications network.

*B. ICT Abatement Potential*

The abatement potential of ICT is a big area that recently attracted a lot of attention. ICT helps organizations and business achieve CO2 emissions reduction and reduced operational and power consumption via digitization of processes and products (e.g. video conferencing, e-commerce, smart-meters), data collection and communications (e.g. eco-driving, forecasting an benchmarking), system integration(e.g. virtual power plants and renewable energy sources integration) and process and functional optimization(automation, control, improved design). The ICT potential in energy efficiency techniques and technologies is also great. On the other side of the spectrum, there is the need for sustainability within the ICT sector. With the current issues and climate change impacts that are evidenced in various ways in temperature increase, more droughts and extreme weather conditions and impact on energy generation. The question of what ICT can do for sustainability becomes more pertinent.

As we have seen, ICT is then an enabler for sustainable business growth and CO2e reductions in other sectors. ICT abatement potential is calculated by GeSi in Figure 2[11]. Energy and transportation sectors are major contributors into GtCO2e [12]. Hence in this paper we argue that there is a great opportunity in the power sector for ICT innovations towards sustainability. In the next section we discuss techniques that help increase power efficiency of ICT and recent progress in wireless mobile cellular networks and fixed broadband networks.

IV. ICT ROLE IN ENERGY EFFICIENCY AND SUSTAINABILITY-ANALYSIS

We have established in the previous section, ICT has a role to play in energy efficiency. ICT has also potential in improving on three pillars of sustainability: economy, environment and social [13,14].

*A. ICT Technologies- Econominc perspective*

Looking at telecommunications networks, the network energy use of ICT will increase by 27% between 2012-2016[15] while the network energy bill will constitute 75% of overall energy bill. This is different in mature and emerging markets as the energy bill as a share of Operational Expenditure (OPEX) will increase from 7-20% between 2012-2016. This dictates the need to act now on energy efficiency measures and to capitalize on ICT abatement potential in process and functional optimization and system integration areas.

Figure 3 highlight the difference in emission growth in communications network between wired and wireless technologies and networks. This is the reason why there is currently a lot of research interest in optimizing wireless networks. Research by green touch[13] showed that right combination of wireless network technology innovations, solutions and specifications could improve energy efficiency by a factor of 1000 by 2015 over 2010 figures.

*B. ICT Technologies - Social perspective*

Social perspective of ICT is manifested in improving communication among members of the society as well contributing to major social and community projects. ICT brings community groups with the same interests and different interests together.

*C. ICT Technologies - enviormental perspective*

The environmental impacts of ICT are not just restricted to Co2 emissions. There are other impacts on air and atmosphere (e.g. generators pollutions powering 2G/3G/4G radio sites), water resources pollution, soil and geology and even cultural heritage. Most of above environmental impacts can be influences of building a radio site equipped with generators and tall tower mast and other equipments. Moreover, population demography segmentation and social-cultural lifestyle of people can also be affected by proliferation of those radio sites and telecomm equipments.

V. SUSTAINABILITY TRENDS IN TELCOMMUNICATIONS

Figure 4 shows the ICT telecommunication network energy gap between the growth of traffic and network operational efficiency. This growing gap is attracting huge interest in innovative technological and architectural solutions.

Several contributions have been made towards tackling this increase e.g. energy transmission protocols design in Long Term Evolution (LTE) mobile networks [16,17] and practical implementations trade-offs of transmission protocols implementations [7,8]. The need for metrics and measurement models is presented with application example in [9]. Recently, the authors have proposed in [7] a hybrid wireless-fibre broadband hybrid Wireless-fiber broadband access system. Current trends include utilizing smart metering technologies and equipments to adjust power load according to changing energy prices (for buildings, businesses, corporate,etc) and self organizing networks concepts for automatic shutdown of radio equipments in radio sites offering data and voice mobile communications services. Those trends will continue along with advances in power amplifier efficiency for mobile base stations, active cooling for data centers and radio sites and fundamental changes in mobile and fixed broadband network architectures.

## VI. CONCLUSIONS

An important feature of ICT is its ability to save energy in other industries by directly reducing operational power consumption within the technical operation of ICT function or indirectly via streamlining processes and cost savings in logistics and capital expenditure. In this paper we showed how ICT addresses sustainability pillars: economy, social and environmental while highlighting areas of recurrent research in enabling technologies such as smart measurements and monitoring of energy consumption and energy management techniques for telecommunications networks. System integration, digitization of products and services and data collection and analysis are three core approaches for enabling huge savings from ICTs across different industries.

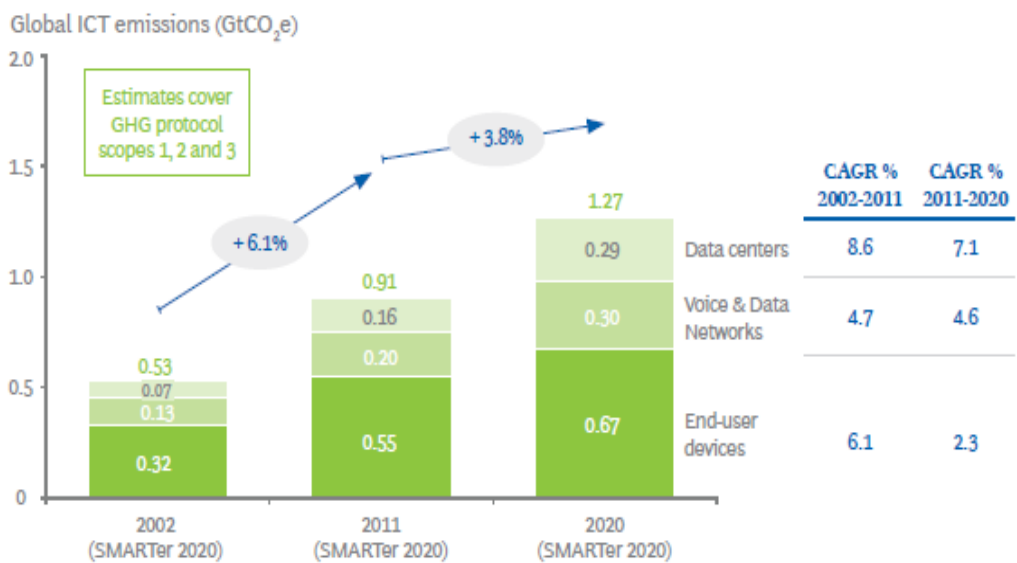

Figure 1: ICT emissions growth from 2002 -2020. [GeSI [11]]

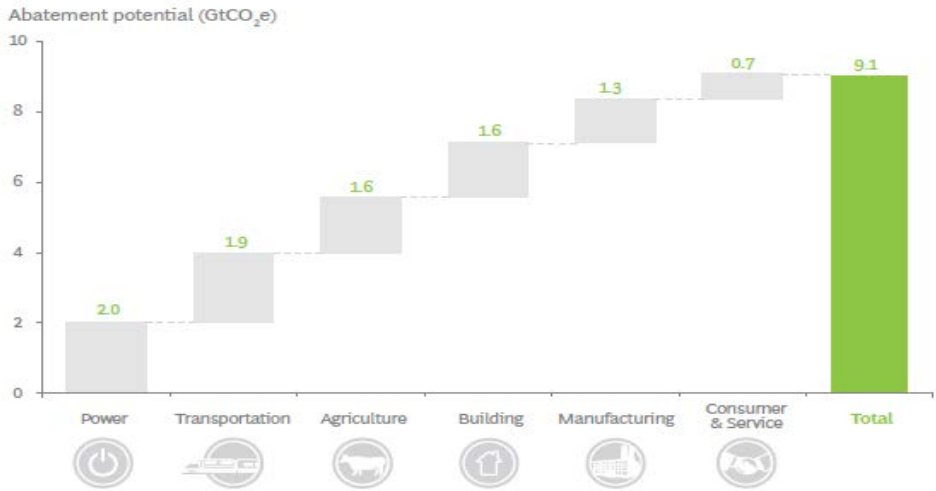

Figure 2: ICT abatement potential by end-user sector in year 2020 [GeSI[11]].

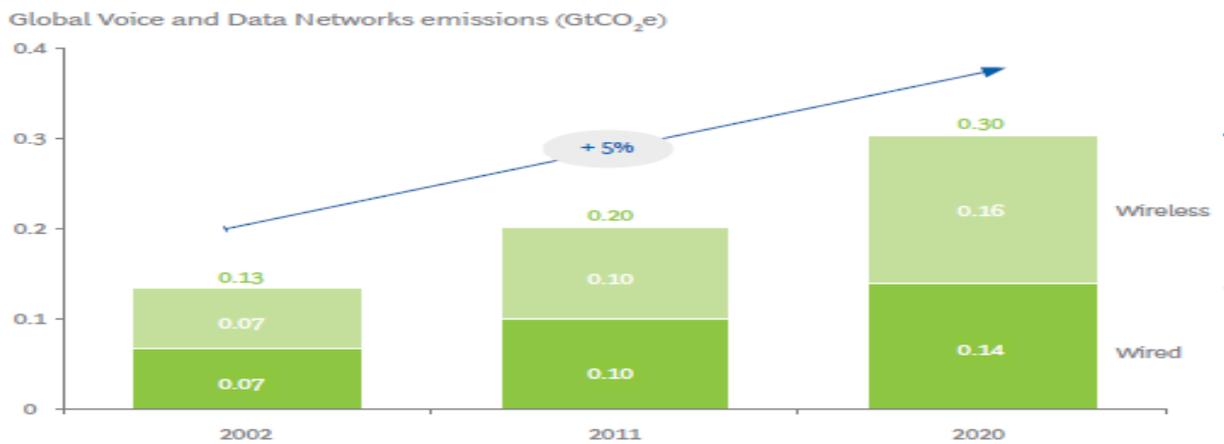

Figure 3: Wired and Wireless communications networks emissions growth from 2002 -2020[GeSI[11]]

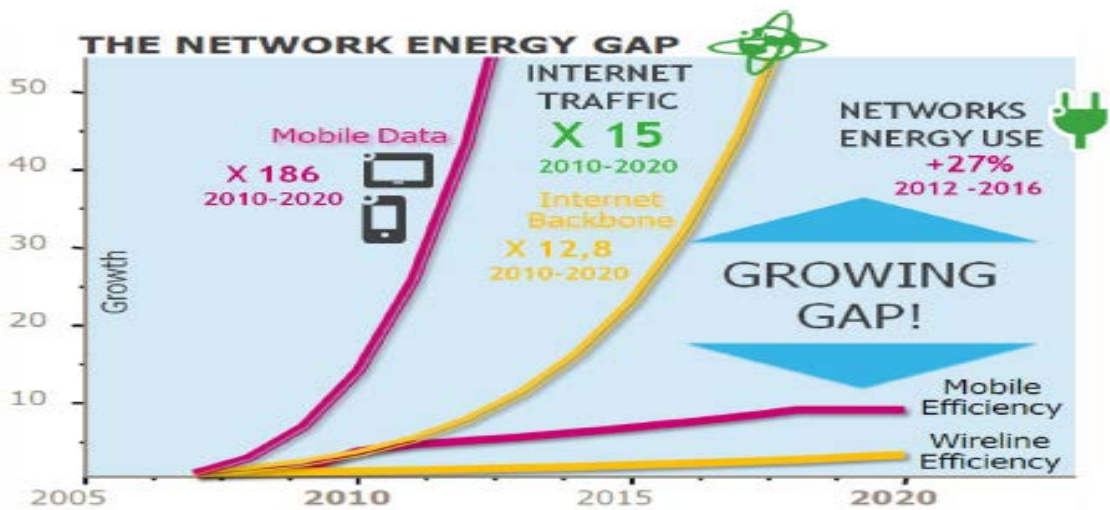

Figure 4: The Network Energy gap between traffic growth and energy consumption [ Green touch[15]]